 \documentclass[final,5p,times,twocolumn,authoryear]{elsarticle}

\usepackage{amssymb}
\usepackage{lipsum}
\usepackage{amsmath}
\usepackage{graphicx,xcolor}
\journal{Nuclear Physics B}

\begin{document}

\begin{frontmatter}



\title{Enhanced water Cherenkov detector for soil moisture detection}


\author[UIS,Udenar]{J. Betancourt}
\author[UIS,UNAB]{C. Sarmiento-Cano}
\author[CAB,CONICET]{I. Sidelnik}
\author[Piensas]{H. Asorey}
\author[UIS]{Y. Domínguez-Ballesteros}
\author[UIS]{L. Miranda-Leuro}
\author[UIS,ULA]{Luis A. Núñez}

\affiliation[UIS]{
            organization={Escuela de Física, Universidad Industrial de Santander},%Department and Organization 
            city={Bucaramanga},
            postcode={680002},
            country={Colombia}}
\affiliation[Udenar]{
            organization={Departamento de Física, Universidad de Nariño},%Department and Organization 
            city={Pasto},
            postcode={520006},
            country={Colombia}}
\affiliation[UNAB]{organization={Departamento de Ciencias Básicas, Universidad Autónoma de Bucaramanga},%Department and Organization
            city={Bucaramanga},
            postcode={680003}, 
            country={Colombia}}
\affiliation[CAB]{organization={Departamento de Física de Neutrones,  Centro Atómico Bariloche},%Department and Organization
            city={San Carlos de Bariloche},
            postcode={R8402AGP}, 
            country={Argentina}}
\affiliation[CONICET]{organization={Consejo Nacional de Investigaciones Científicas y Técnicas (CONICET)},%Department and Organization
            city={Buenos Aires},
            %postcode={R8402AGP}, 
            country={Argentina}}
\affiliation[Piensas]{
            organization={piensas.xyz, Centro Tripark Las Rozas},%Department and Organization 
            city={Madrid},
            postcode={28232},
            country={Spain}}
\affiliation[ULA]{
            organization={Departamento de Física, Universidad de Los Andes},%Department and Organization 
            city={Mérida},
            postcode={5101},
            country={Venezuela}}            
\begin{abstract}
This work evaluates the ability of a water Cherenkov detector to measure thermal neutrons and explores its application to soil-moisture monitoring. We study a NaCl-doped detector and model its response to (i) monochromatic thermal neutrons and (ii) the natural thermal-neutron flux expected from dry soil at the elevation of Bucaramanga, Colombia. The ambient flux is estimated with the ARTI framework, and the detector response is simulated with MEIGA in Geant4. Doping with NaCl introduces additional capture channels on $^{35,37}\mathrm{Cl}$ and $^{23}\mathrm{Na}$; in particular, $^{35}\mathrm{Cl}$ has a thermal-neutron absorption cross section up to two orders of magnitude larger than hydrogen, boosting the capture signal. Our results indicate that water Cherenkov detectors can detect thermal neutrons with practical sensitivity under field conditions, enabling their integration into precision agriculture networks for soil moisture sensing. More broadly, this approach extends the cosmic-ray detection range of Cherenkov detectors using non-toxic, low-cost materials.
\end{abstract}



\begin{keyword}
Cherenkov detector \sep thermal neutrons \sep soil moisture \sep Geant4 simulation \sep precision agriculture \sep cosmic rays 


\end{keyword}

\end{frontmatter}

%\tableofcontents



\section{Introduction}
\label{introduction}
Soil moisture plays a key role in the hydrological cycle, governing important processes such as rainfall infiltration, surface runoff, plant water uptake, and groundwater recharge. Reliable estimates of soil moisture are fundamental for a wide range of applications, including agriculture, water resource management, and climate modelling. Nevertheless, capturing soil moisture at ecologically and hydrologically relevant spatial scales remains challenging. Conventional in situ methods (such as time-domain reflectometry probes or gravimetric sampling) yield only point-scale measurements, typically representing areas of just a few centimetres. In contrast, satellite remote sensing provides spatially averaged estimates over footprints that span several kilometres. This stark mismatch in spatial resolution introduces a critical observational gap: point measurements cannot resolve field-scale heterogeneities, while satellite data often lack the necessary resolution to inform local-scale hydrological processes (see \cite{FlynnWyattMcInnes2021} and references therein).

In response to this limitation, cosmic-ray neutron sensing (CRNS) has emerged as an innovative technique for non-invasive soil moisture monitoring at the field scale. A ground-based CNRS detects fast neutrons ($10~{\rm MeV} \leftrightarrow 1~{\rm GeV}$) produced by interactions of galactic cosmic rays with atmospheric nuclei. These fast neutrons penetrate the soil and undergo energy loss, predominantly through elastic scattering with hydrogen nuclei, which are primarily found in soil water. Because hydrogen is an efficient neutron moderator, the near-surface fast neutron flux is inversely correlated with the soil's hydrogen content. As a result, higher soil moisture corresponds to a lower count rate of fast neutrons above ground. By calibrating neutron count rates against collocated gravimetric soil moisture data, one can derive continuous estimates of volumetric water content over a horizontal footprint of approximately $100$m to $200$m in radius and a depth of up to $50$cm. This approach effectively bridges the gap between point-scale and satellite-based observations, offering valuable, spatially integrated soil moisture data that complement traditional techniques \citep{ZredaEtal2008}.

Most conventional CRNS instruments rely on moderated gas-filled detectors, typically employing proportional counters filled with either helium-3 ($^3{\rm He}$) or boron trifluoride (${\rm BF}_3$). These tubes are embedded in polythene moderators that slow down incident-fast neutrons to thermal energies. Upon thermalisation, the neutrons are captured by the detector gas, producing charged particles that generate electronic pulses, which are then counted as neutron events. Although these systems are effective, $^3{\rm He}$-based detectors are now prohibitively expensive due to a well-documented and ongoing global shortage of helium-3, a rare isotope primarily obtained as a byproduct of nuclear weapons disarmament (see \cite{ZredaEtal2008, ZredaEtal2012, EvansEtal2016}). ${\rm BF}_3$-based detectors, while less costly, involve the use of highly toxic and corrosive gas, raising significant safety and environmental concerns \citep{franzEtal2012, AndreasenEtal2017}. These economic and regulatory limitations have impeded the broader deployment of CRNS technology beyond specialised research settings, thereby driving the development of alternative, low-cost, and safer neutron detection methods.

In response to the limitations of conventional $^3{\rm He}$ and ${\rm BF}_3$-based neutron detectors, several innovative sensor technologies have been developed to either replace or complement traditional CRNS instruments. An example of this type of cosmic-ray neutron sensor integrates inorganic scintillators with plastic scintillator materials and photomultiplier tubes to detect neutron interactions \citep{StevanatoEtal2019}. Another promising alternative is the boron-lined proportional counter, which substitutes the gas-filled design with solid boron-10 layers that capture thermal neutrons via the $^{10}{\rm B}(n,\alpha)^7{\rm Li}$ reaction. This design preserves the fundamental detection principle of $^3{\rm He}$ tubes while leveraging the abundance and non-toxicity of boron, and it also offers the potential for resolving neutron energy spectra \citep{BogenaEtal2021}.  A further advancement involves the use of $^6{\rm Li}$ foil in multi-wire proportional counters. Thermal neutrons are detected through the $^6{\rm Li}(n,\alpha)^3{\rm H}$ reaction, which releases energetic charged particles (alpha and triton) that generate a measurable signal. The widespread availability of lithium, especially from the battery industry, makes these sensors significantly more cost-effective, reducing unit costs to approximately half of those associated with $^3{\rm He}$ detectors \citep{AndreasenEtal2017, StowellEtal2021}. Collectively, these developments, ranging from scintillator-based systems to boron- and lithium-enhanced counters, represent a significant step forward in reducing costs and expanding access to cosmic-ray neutron soil moisture sensing.

In addition to conventional tube-and-scintillator-based detectors, researchers have investigated large-volume Water Cherenkov Detectors (WCDs) as a promising alternative for environmental neutron sensing. Water Cherenkov Detectors, essentially tanks filled with water and equipped with photomultiplier tubes (PMTs), are well-established in astroparticle physics for detecting relativistic charged particles, via the Cherenkov radiation they emit while traversing the water medium. Although pure water exhibits some sensitivity to fast neutrons through elastic scattering with protons and subsequent neutron capture on hydrogen, the intrinsic detection efficiency remains modest, typically on the order of a few tens of per cent \citep{SidelnikEtal2017, SidelnikEtal2020A}. Recent work has demonstrated that doping the water with certain salts, notably sodium chloride (NaCl), significantly enhances neutron detection performance. The key mechanism involves thermalised neutrons capturing on chlorine nuclei, which subsequently emit a cascade of high-energy gamma. These gamma interact with the water to produce secondary electrons, which generate Cherenkov light and thereby amplify the observable signal~\citep{SidelnikEtal2020B}.

Experimental studies using a $1 m^3$ prototype WCD revealed a marked improvement in fast neutron detection efficiency, from approximately $19$\% in pure water to about $44$\% with a $2.5$\% NaCl solution. Monte Carlo simulations conducted using the Geant4 toolkit corroborated these findings. They provided a detailed understanding of the physical processes involved, confirming the enhanced Cherenkov photon yield resulting from neutron capture-induced gamma cascades. Importantly, both water and NaCl are inexpensive, non-toxic, and widely available, positioning this technology as a low-cost and scalable option for field-deployable neutron monitoring systems. In this work, we reinforce the idea that salt-doped WCDs could provide a viable complement, or even an alternative, to traditional $^3{\rm He}$-based neutron monitors in applications ranging from environmental radiation surveillance to the detection of special nuclear materials~\citep{SidelnikEtal2020C}.

Building upon recent advancements in cosmic-ray neutron sensing, this study investigates the feasibility of using a sodium chloride-doped Water Cherenkov Detector to measure soil moisture at the field scale. We present the design of a WCD specifically optimised for neutron detection and evaluate its performance through comprehensive Monte Carlo simulations of cosmic-ray interactions with the atmosphere and the terrestrial environment. Particular emphasis is placed on analysing the detector's sensitivity to variations in soil water content, which modulate the flux of incoming cosmic-ray-induced neutrons via hydrogen-rich attenuation processes. By validating the detector response using Geant4-based simulations and correlating the neutron count rates with modelled soil moisture changes, we demonstrate that a salted WCD can effectively resolve hydrologically relevant variations in near-surface water content. The results underscore the potential of this technology as a cost-effective, scalable, and non-invasive tool for continuous soil moisture monitoring, thus expanding the current repertoire of hydrological measurement techniques.

This paper is organised as follows. Section \ref{CosmicNeutronFlux} describes the simulation of cosmic-ray-induced neutron fluxes at ground level, emphasising the modelling of atmospheric interactions and the influence of soil moisture on the neutron energy spectrum. Section \ref{EnhancedWCD} introduces the enhanced Water Cherenkov Detector design, details the implementation of Geant4-based simulations using the MEIGA framework, and presents the methodology for assessing neutron detection efficiency under varying salinity conditions. Section \ref{ResultsDiscussion} presents the main results and discusses the impact of sodium chloride doping on the detector's response, including capture efficiency, Cherenkov photon yield, and signal stability under realistic neutron backgrounds. Finally, Section \ref{ResultsDiscussion} summarises the principal conclusions and outlines future directions for experimental validation and field deployment.

\section{Cosmic neutron flux}
\label{CosmicNeutronFlux}
\subsection{Atmospheric Neutron simulations}
A reliable estimation of the cosmic neutron flux at ground level requires accurately modelling the entire energy spectrum of secondary particles produced by the interaction of galactic cosmic rays with the Earth's atmosphere. This task presents two significant challenges. First, it involves capturing the development of high-energy particle cascades initiated in the upper atmosphere. Second, it requires modelling the transport and moderation of low-energy neutrons in the lower atmosphere, particularly within the last $2000$~m above ground level, where their interactions significantly shape the thermal and epithermal components of the neutron flux.

\textsc{FLUKA} (for \emph{FLUktuierende KAskade}\footnote{{\tt https://fluka.cern/about}}) is a general-purpose Monte-Carlo framework that tracks neutron histories continuously from the GeV domain down to thermal energies of $10^{-5}\text{eV}$. This code calculates energy, radial distance, depth, and arrival time on user-defined atmospheric layers, thereby resolving the millisecond-delayed thermal cloud that diffuses hundreds of metres from the shower axis. This end-to-end treatment links ground-level neutron observables to the hadronic energy budget, enabling clear discrimination among proton, iron, and photon primaries without the need for auxiliary parameterisations. No other public code delivers an equally self-consistent description across nine orders of magnitude in neutron energy, while maintaining the extensive experimental validation that \textsc{FLUKA} has accumulated over several decades. The main constraints remain operational rather than physical: users must construct detailed geometries, select suitable cross-section libraries, and tune numerical thresholds, which lengthens the setup phase (see \cite{SchimassekEtal2024} and the references therein). 

On the other hand, recent studies have reported the use of \textsc{URANOS} (Ultra-Rapid Neutron-Only Simulation) to trace cosmic-ray neutrons for hydrological applications. The code, conceptually derived from \textsc{Geant4}, accelerates calculations by propagating neutrons exclusively, an approach that sacrifices several physical couplings for \emph{thermal} neutron soil-moisture retrievals~\citep{KohliEtal2023}. Because the framework ignores charged and electromagnetic secondaries, it cannot internally adapt the low-energy neutron flux to atmospheric pressure, geomagnetic latitude, or vegetation water equivalent. Users must externally prescribe the complete source spectrum. Furthermore, the bundled cross-section libraries accurately resolve epithermal interactions but treat the complex resonance structure governing thermal moderation more coarsely. These simplifications can bias absolute soil-moisture estimates when hydrogen pools in litter, intercepted rainfall, or biomass dominate the slowing-down cascade~\citep{Dominguezballesteros2023}.

In our case, we address the challenges mentioned above by tracing neutron histories using the ARTI toolkit \citep{SarmientocanoEtal2022}. This framework, developed by the Latin American Giant Observatory (LAGO)\footnote{\texttt{https://lagoproject.net/}}, is a modular simulation suite designed to quantify the atmospheric background of secondary cosmic-ray particles. It combines three core components: Magnetocosmics \citep{desorgher_magnetoscosmics_2003}, which calculates the geomagnetic rigidity cutoff; CORSIKA~\citep{dheck_corsika_1998}, which simulates the propagation and interaction of high-energy primaries in the atmosphere; and Geant4~\citep{AgostinelliEtal2003}, a widely adopted package for modelling the detector's response to radiation.

Specifically, we used Geant4 to simulate the transport and interactions of low-energy neutrons within the bottom $2000$~m of the atmosphere~\citep{ALLISON2016186}. The atmospheric model used for this simulation assumed an ideal gas in hydrostatic equilibrium, characterised by a stratified density profile that decreases with altitude. The composition of the atmosphere, primarily consisting of nitrogen, oxygen, and argon, adhered to the standard proportions specified in the U.S. Standard Atmosphere\footnote{\texttt{https://www.pdas.com/atmos.html}}. The simulation domain is divided into ten horizontal layers of equal thickness. Each is assigned a constant density. We employed the \texttt{QGSP\_BERT\_HP} physics list to describe particle interactions, including high-precision models for elastic scattering, absorption, and thermal moderation of neutrons.

By combining ARTI's capability for simulating high-energy particle cascades with Geant4's strength in modelling low-energy neutron transport, we obtained a complete and consistent reconstruction of the cosmic neutron spectrum at ground level. In particular, we supplemented ARTI's output by performing Geant4-based simulations to account for neutrons with momenta below $200$ MeV/c, which CORSIKA does not propagate. This allowed us to track the propagation of these low-energy neutrons through the final layers of the atmosphere.

%%%%%%%%%%%%%%
\begin{figure}[h]
\centering
\includegraphics[width=3in]{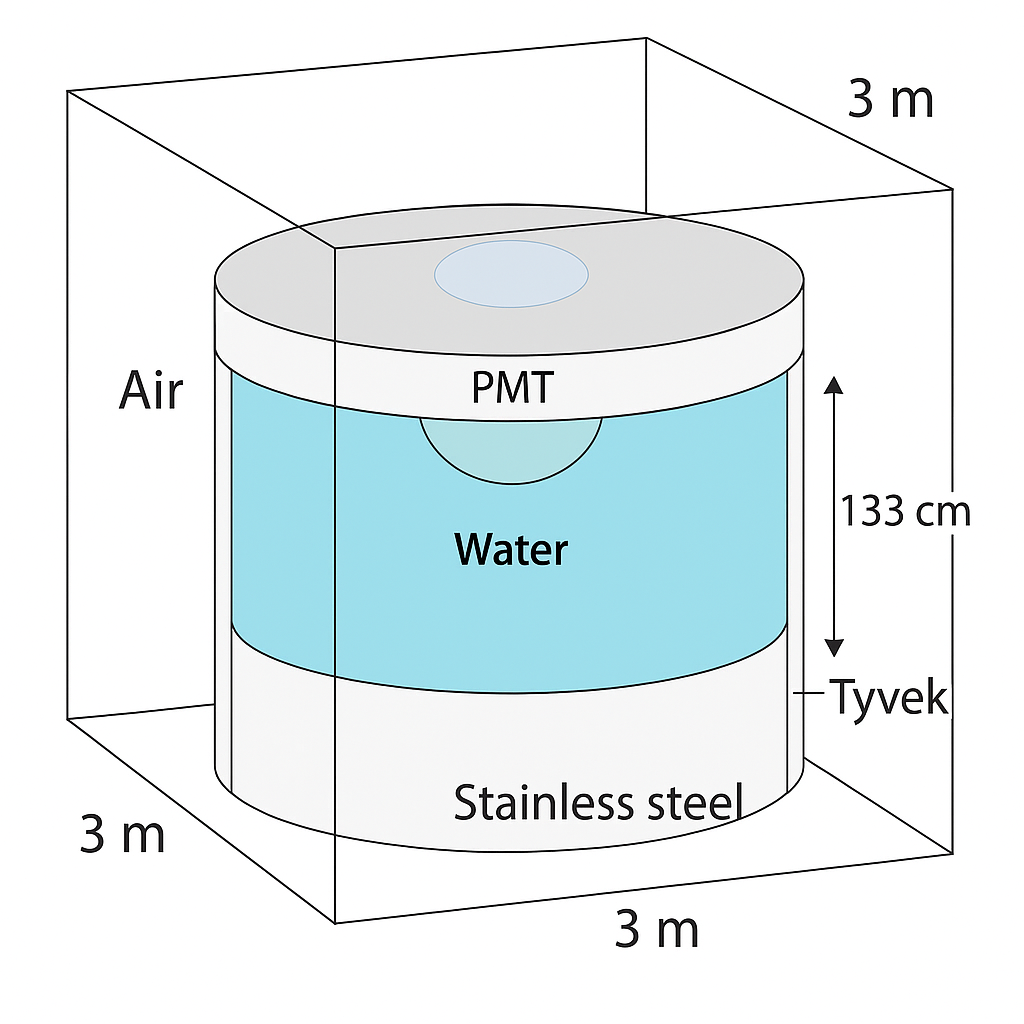}
\caption{A water Cherenkov detector tank, as well as its variant with a water and sodium chloride mixture, was placed inside a cubic room filled with air, measuring $3 \times 3 \times 3$ meters. The tank has a cylindrical geometry and features a hemispherical structure on top representing the photomultiplier tube.}
\label{sim-tank}
\end{figure}

\subsection{Neutron Flux at ground level}
\label{subsec:neutron-flux}
%%%%%%%%%
We implemented a dedicated \textsc{Geant4} module to model neutron moderation and albedo at the soil-air interface. Thus, recovering the thermal tail of the cosmic-ray neutron spectrum that dominates the detector response and varies strongly with soil moisture. We surrounded a $1$m-deep rectangular block of soil with a world volume filled with dry \texttt{G4\_AIR}, which allowed neutrons to propagate seamlessly from the atmosphere into the ground and back into the air.

We specified the soil composition based on the representative values for Colombian dry soils reported by \cite{Juarezsanz2006}: Oxygen $49$\%, silicon $33$\%, aluminium $7.1$\%, and smaller fractions of iron, calcium, potassium, hydrogen, and trace elements. 

The extended framework tracks neutron trajectories and energy losses inside both the atmosphere and the soil, quantifies the albedo contribution emerging from the ground, and produces moisture-dependent thermal and epithermal spectra at the surface. These spectra serve as indispensable inputs for interpreting the response of NaCl-doped water Cherenkov detectors and for distinguishing between environmental effects and intrinsic detector performance.

\begin{figure}
    \centering
    \includegraphics[width=0.99\linewidth]{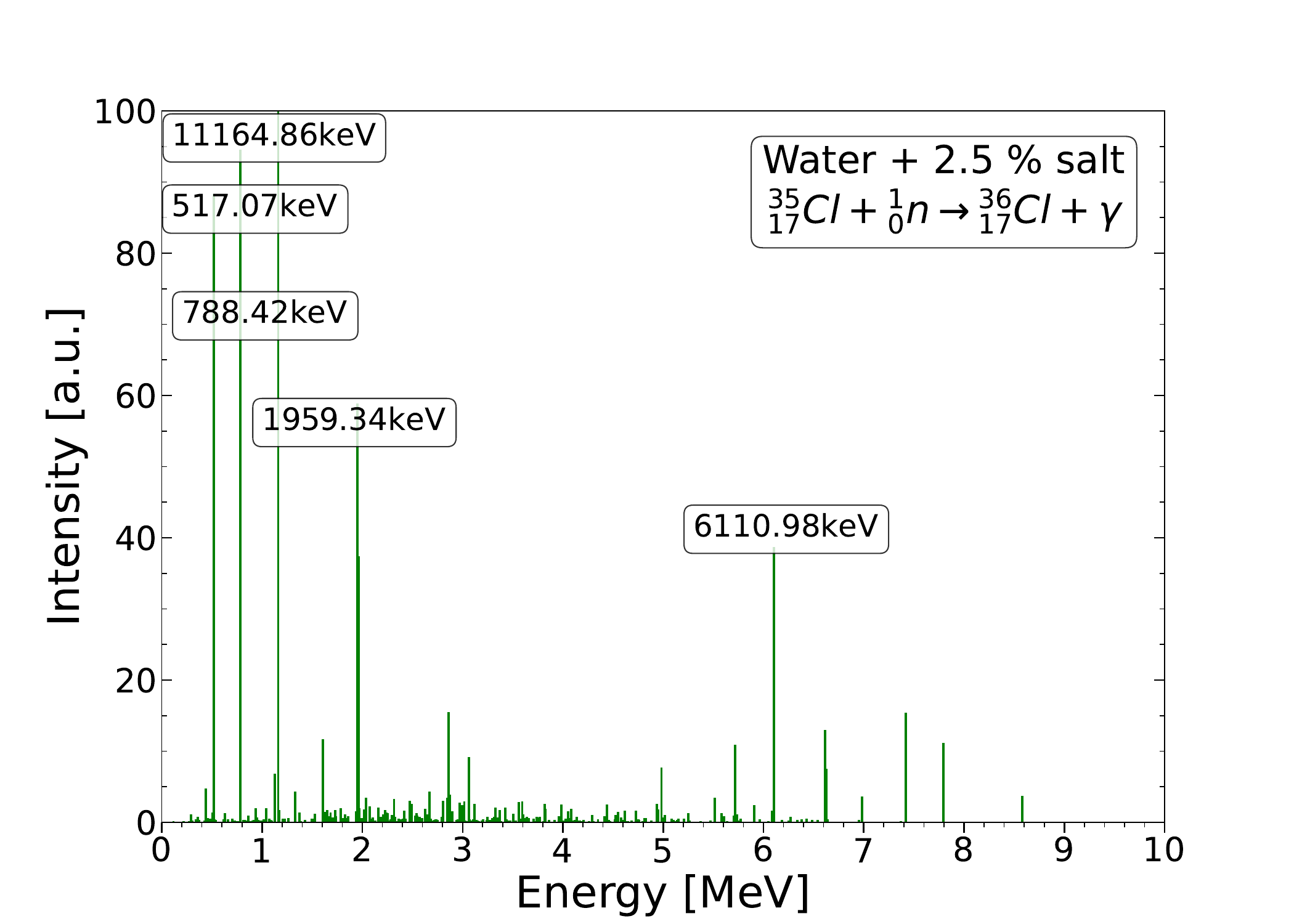}    
    \includegraphics[width=0.99\linewidth]{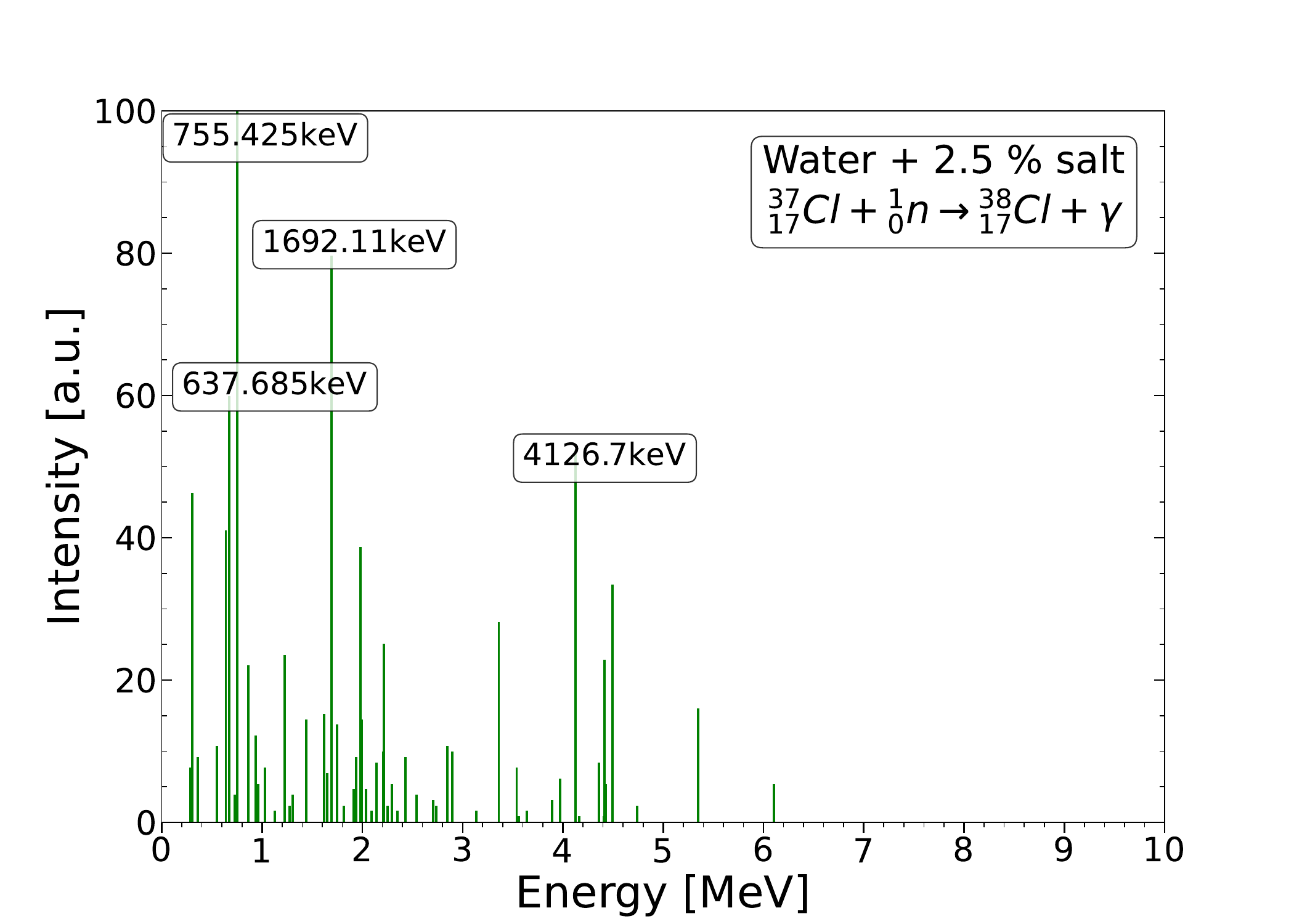}
    \caption{Spectral lines corresponding to the gamma spectrum produced as a result of neutron capture from the GEANT 4 simulation. The contributions from $^{35}_{17}Cl$ (top) and $^{37}_{17}Cl$ (bottom) in a solution of water with 2.5 $\%$ salt are shown.}
    \label{fig-lineas-esp}
\end{figure}

\section{Enhanced Water Cherenkov Detector simulations}
\label{EnhancedWCD}
We investigated the interaction of thermal neutrons (those with energies between $1$~meV and $1$~keV) within the active volume of a cylindrical WCD. The study compared two media: pure water and water doped with sodium chloride at mass fractions of $2.5$\%, $5$\%, and $10$\%.

\subsection{Water Cherenkov Detector Model}
We performed Monte Carlo simulations with the MEIGA framework~\citep{TaboadaEtal2022} to quantify the response of a water Cherenkov detector to thermal neutrons from cosmic-ray showers. MEIGA, created for muography, embeds the Geant4 11.1 transport engine and exposes an object-oriented interface, allowing users to prescribe geometry, material composition, optical properties, and readout parameters in a single configuration file. The framework propagates each ground-level particle through the model, tracks wavelength-dependent Cherenkov photons, and reconstructs digitised signals, thereby supporting rigorous efficiency studies and detector optimisation.
\begin{figure}
    \centering
    \includegraphics[width=1.0\linewidth]{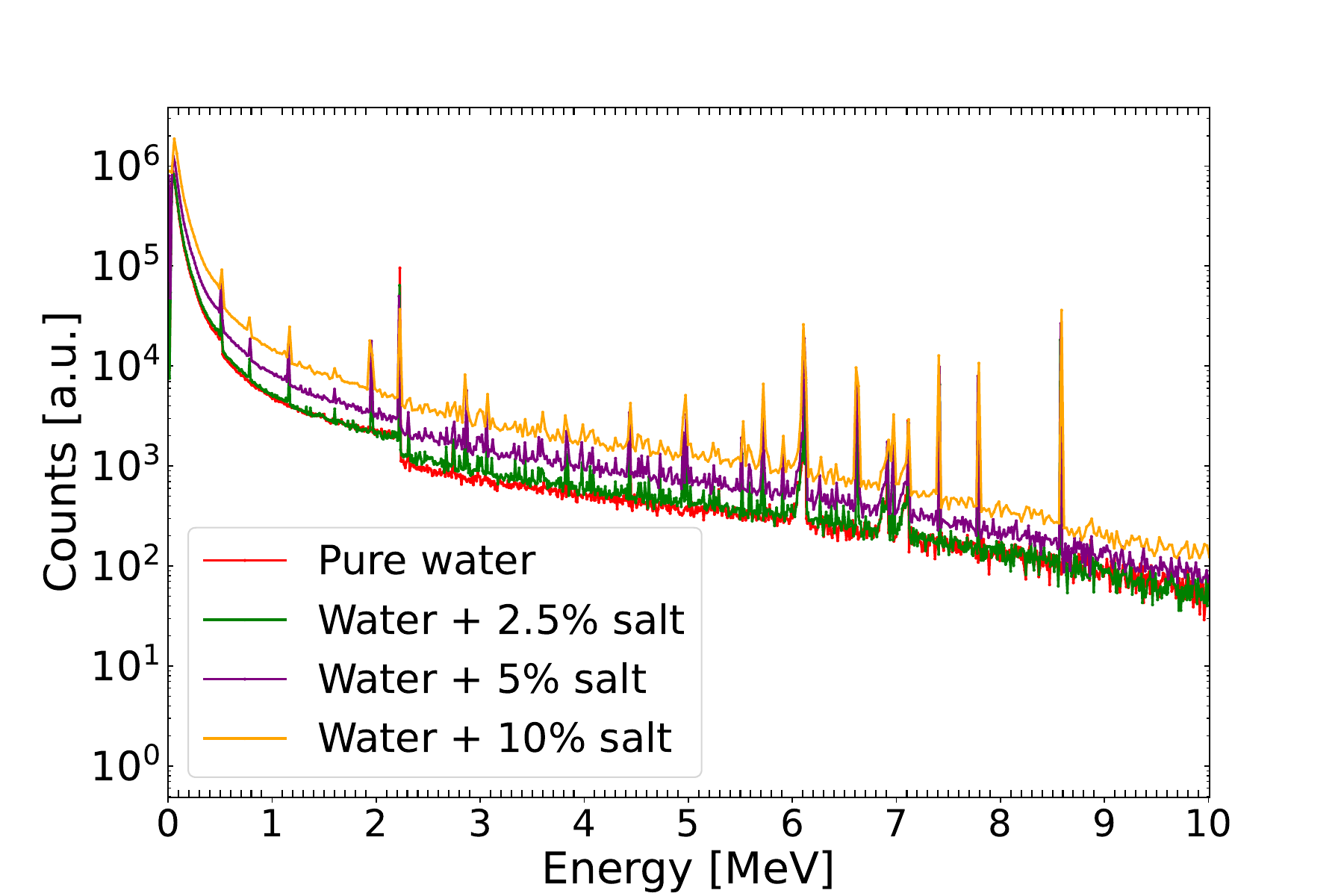}
    \includegraphics[width=0.48\textwidth]{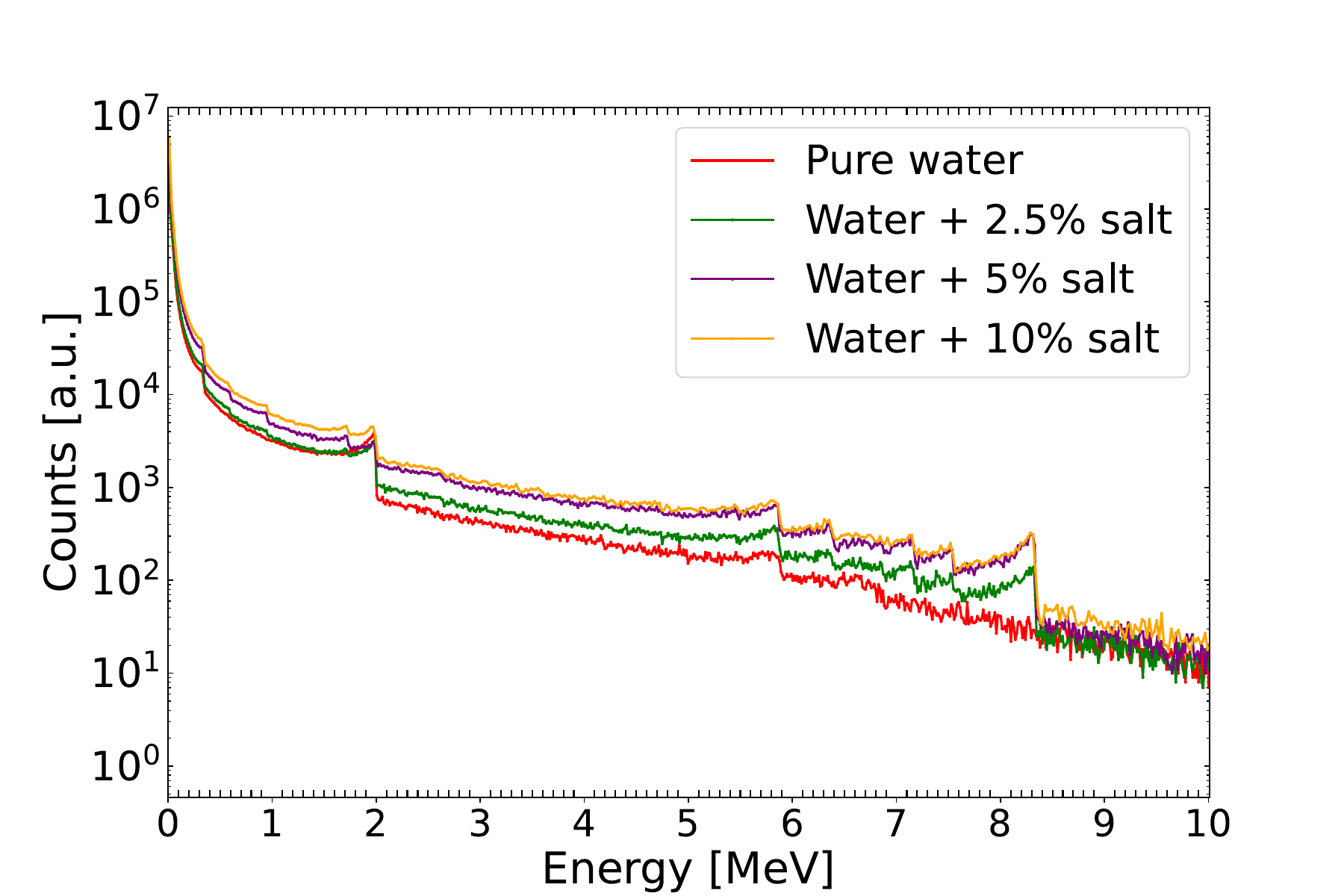}
    \caption{\textbf{Top:} Differential energy spectrum of the $\gamma$ cascade generated inside the WCD by a secondary neutron spectrum coming from an ARTI simulation\cite{SarmientocanoEtal2022}.  Curve colours denote the detection medium: pure water (red), $\text{H}_{2}\text{O}+2.5\,\%\ \text{NaCl}$ (green), $5\,\%$ (purple), and $10\,\%$ (orange).  \textbf{Bottom:} Corresponding energy spectrum of secondary electrons produced by Compton scattering and pair production of the $\gamma$ population from the left figure. The progressive shift toward higher yields at larger salt fractions reflects the enhanced $\mathrm{^{35}Cl}(n,\gamma)$ capture rate and the ensuing increase in high-energy $\gamma$ emission.}
    \label{fig:GamaElec-esp}
\end{figure}

We translate the design shown in Fig.~\ref{sim-tank} into a Geant4 geometry reproducing a cylindrical tank, a diffuse titanium dioxide liner, a ten-inch photomultiplier tube, and an active volume of $1$~m$^{3}$. The simulation considers two detection media: ultra-pure water and aqueous NaCl solutions with mass fractions of $2.5$\%, $5$\%, and $10$\%. The $^{35}\mathrm{Cl}(n,\gamma)$ capture reaction releases an $\sim8.6$~MeV $\gamma$ cascade that drives Compton electrons and, hence, amplifies the Cherenkov yield in proportion to the salt concentration. This model furnishes a self-consistent platform for evaluating the gain in neutron-tagging efficiency that NaCl doping provides under realistic optical and electronic conditions~\citep{SidelnikEtal2020A}.

Figure~\ref{fig-lineas-esp} shows the Geant4 output gamma spectra from neutron capture reactions on $^{35}\mathrm{Cl}$ (top) and $^{37}\mathrm{Cl}$ (bottom). It is important to note that these gamma emissions must be weighted by the natural isotopic abundances of chlorine: approximately 75\% for $^{35}\mathrm{Cl}$ and 24\% for $^{37}\mathrm{Cl}$ when accounting for the final gamma production coming from neutron capture. Furthermore, since the neutron capture cross section for $^{35}\mathrm{Cl}$ is two orders of magnitude larger than that for $^{37}\mathrm{Cl}$, most of the neutron capture signal is expected to originate from the gamma lines associated with $^{35}\mathrm{Cl}$, as shown in the left panel of Fig.~\ref{fig-lineas-esp}.

\subsection{Simulation strategy}
The simulation strategy proceeded in three steps. First, we validated our MEIGA implementation by replicating the experimental and numerical results reported in \cite{SidelnikEtal2020C}. This guarantee the consistency in detector geometry, material definitions, and physical processes, particularly for neutron interactions and Cherenkov photon production. 

Second, we systematically assessed the detector's sensitivity across the thermal energy range by simulating its response to monochromatic neutrons with energies between $1$~meV and $1$~keV. For each energy value, we injected neutrons isotropically into the detector and recorded the resulting Cherenkov photon yield and interaction channels. This allowed us to characterise the energy-dependent efficiency of neutron detection for both water and salt-doped configurations.

Finally, we simulated the detector response under continuous exposure to a realistic cosmic thermal neutron flux. This flux was derived from a Geant4-based soil model representative of dry conditions at the altitude of Bucaramanga and integrated over a 12-hour exposure period. The simulation captured the cumulative signal generated by ambient albedo neutrons, enabling a comparative analysis of detection rates with and without NaCl doping.

\subsection{Validation of the detector simulation framework}
We began by benchmarking the MEIGA \& Geant4 chain against the secondary neutron spectrum produced by an ARTI simulation, as can be seen in \cite{SarmientocanoEtal2022},  suited for Bucaramanga. We injected these neutrons from a point source placed $2$~cm outside the WCD wall on the detector's symmetry axis, thereby reproducing the reference geometry with accuracy.

After entering the active volume, neutrons undergo successive elastic and inelastic collisions that moderate their energy to the thermal regime. Thermalised neutrons subsequently capture by the nuclei available in the medium. In NaCl doped water, the principal capture targets are $^{35}\mathrm{Cl}$, $^{37}\mathrm{Cl}$ and $^{23}\mathrm{Na}$. In pure water, the dominant channel involves Hydrogen, $\mathrm{H}$, with a smaller contribution from Oxygen, $\mathrm{O}$. Figure~\ref{fig-lineas-esp} displays the simulated capture-line spectrum and reproduces the hypothesis from \cite{SidelnikEtal2020C}, thus validating the NaCl-doped Water Cherenkov detector configuration.

Each capture event releases a cascade of $\gamma$ (Fig.~\ref{fig:GamaElec-esp} top). Compton scattering and pair production convert these photons into secondary electrons, whose energy spectrum is shown in the bottom panel of Figure~\ref{fig:GamaElec-esp}. Electrons that exceed the Cherenkov threshold emit prompt light pulses that propagate through the water. The Tyvek liner, bulk attenuation length, and refractive indices determine the photon transport, while the PMT wavelength-dependent quantum efficiency governs the photon-to-photoelectron conversion. 

To compare the gamma production from neutron capture, we analysed the difference between the gamma spectra of the various NaCl mixtures and that of pure water, as shown in Fig.\ref{fig:GamaElec-esp} (top). 
As a first approximation, one can observe distinct peaks around 6-7~MeV, which correspond to high-energy gammas produced by the chlorine content in the solution. Conversely, the 2.22~MeV peak serves as an indicator of the competition between hydrogen and chlorine in neutron capture. In the pure water case, the gamma emission is dominated by this 2.22~MeV line; however, as the chlorine concentration in the solution increases, the intensity of this peak decreases by approximately 33\%, 48\%, and 61\% for 2.5\%, 5\%, and 10\% NaCl concentrations, respectively. This trend highlights the overall increase in gamma emission from chlorine lines relative to hydrogen.

\begin{figure}[t]
\centering
\includegraphics[width=0.48\textwidth]{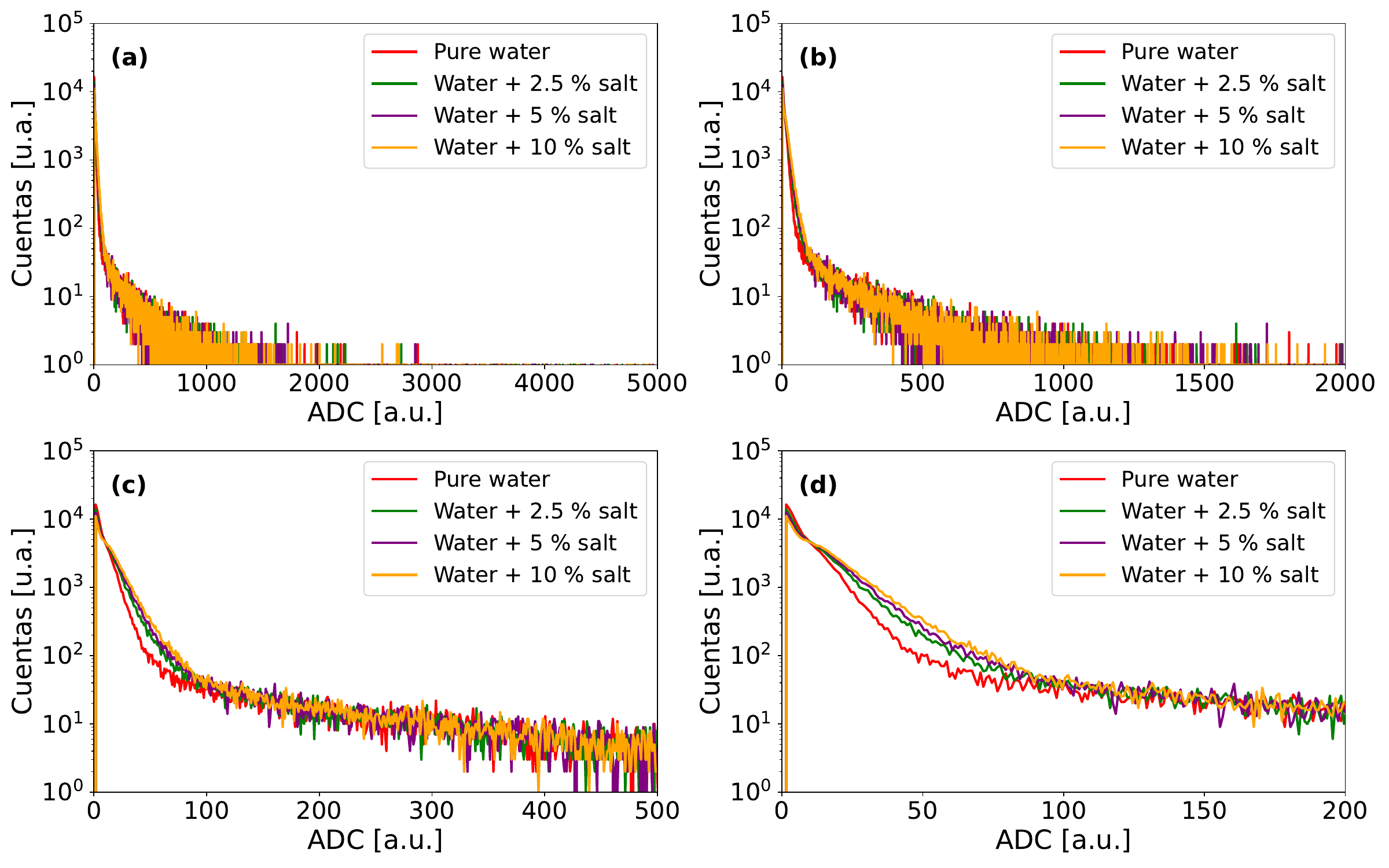}
\caption{Number of Cherenkov photons detected by the PMT, taking into account its quantum efficiency, after being emitted by electrons within the detector volume produced by the electron spectra that can be seen in Fig. \ref{fig:GamaElec-esp} right.}
\label{Num-pho}
%\end{figure}
\end{figure}

Figure \ref{Num-pho} shows the charge histograms obtained with the WCD for four detector volume configurations: pure water (\textcolor{red}{red}), water with 2.5\% salt (\textcolor{green}{green}), water with 5\% salt (\textcolor{purple}{purple}), and water with 10\% salt (\textcolor{orange}{orange}). Panel (a) presents the histogram over the full ADC range, while panels (b), (c), and (d) correspond to zoomed views in the intervals 0--2000, 0--500, and 0--200 ADC, respectively. In the latter range, the slope variation among the different detector configurations becomes more evident. For 0--15 ADC, the dominant contribution arises from pure water, associated with low-energy events. In the interval 15--100 ADC, the saline solutions contribute more significantly, with the 10\% salt configuration exhibiting the highest contribution in this intermediate energy range, where it is expected that the neutrons are dominant. All this as reported in \cite{SidelnikEtal2020A,SidelnikEtal2020C}. 

\begin{figure}[t]
\centering
        \vspace*{-0.5cm}
\includegraphics[width=0.48\textwidth]{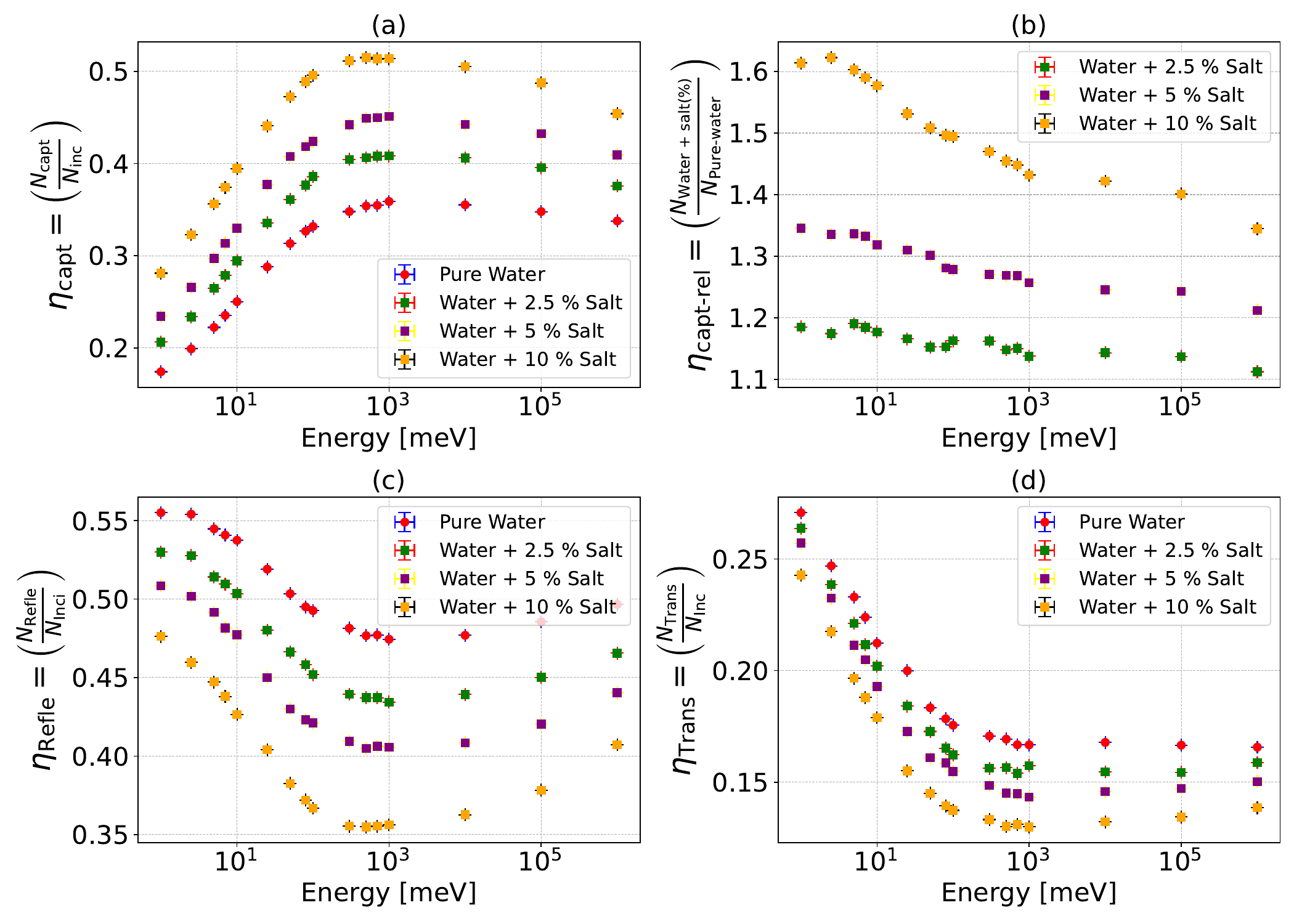}
\caption{Detector response to $10^{5}$ vertically incident mono-energetic neutrons
($1~\mathrm{meV} \le E_{n} \le 1 ~\mathrm{keV}$). Curves compare four target media: pure water (\textcolor{red}{red}), water + $2.5$ \% NaCl (\textcolor{green}{green}), water + $5$ \% NaCl (\textcolor{purple}{purple}), and water + $10$ \% NaCl (\textcolor{orange}{orange}). \textbf{(a)~Capture efficiency} $\eta_{capt}$. NaCl concentration raises the capture probability linearly at all energies and doubles it near $E_{n} \simeq 100~\mathrm{meV}$. \textbf{(b)~Relative gain} $\eta_{capt,rel}= \eta_{capt}(salt)/\eta_{capt}(water)$. Even $2.5$ \% salt delivers a $\sim 15$\% boost; gains saturate beyond a $5$ \% loading. \textbf{(c)~Reflection coefficient} $\eta_{Refle}$. Salted solutions suppress neutron back-scattering; $10$ \% NaCl cuts $\eta_{Refle}$ by an order of magnitude in the thermal region. \textbf{(d)~Transmission coefficient $\eta_{Trans}$.} Higher salt fractions curtail lateral escape; a $10$ \% mixture confines virtually all thermal neutrons within the detector volume.}
\label{cap-det-term}
\end{figure}

\begin{figure}[h]
\centering
\begin{center}
\includegraphics[width=0.48\textwidth]{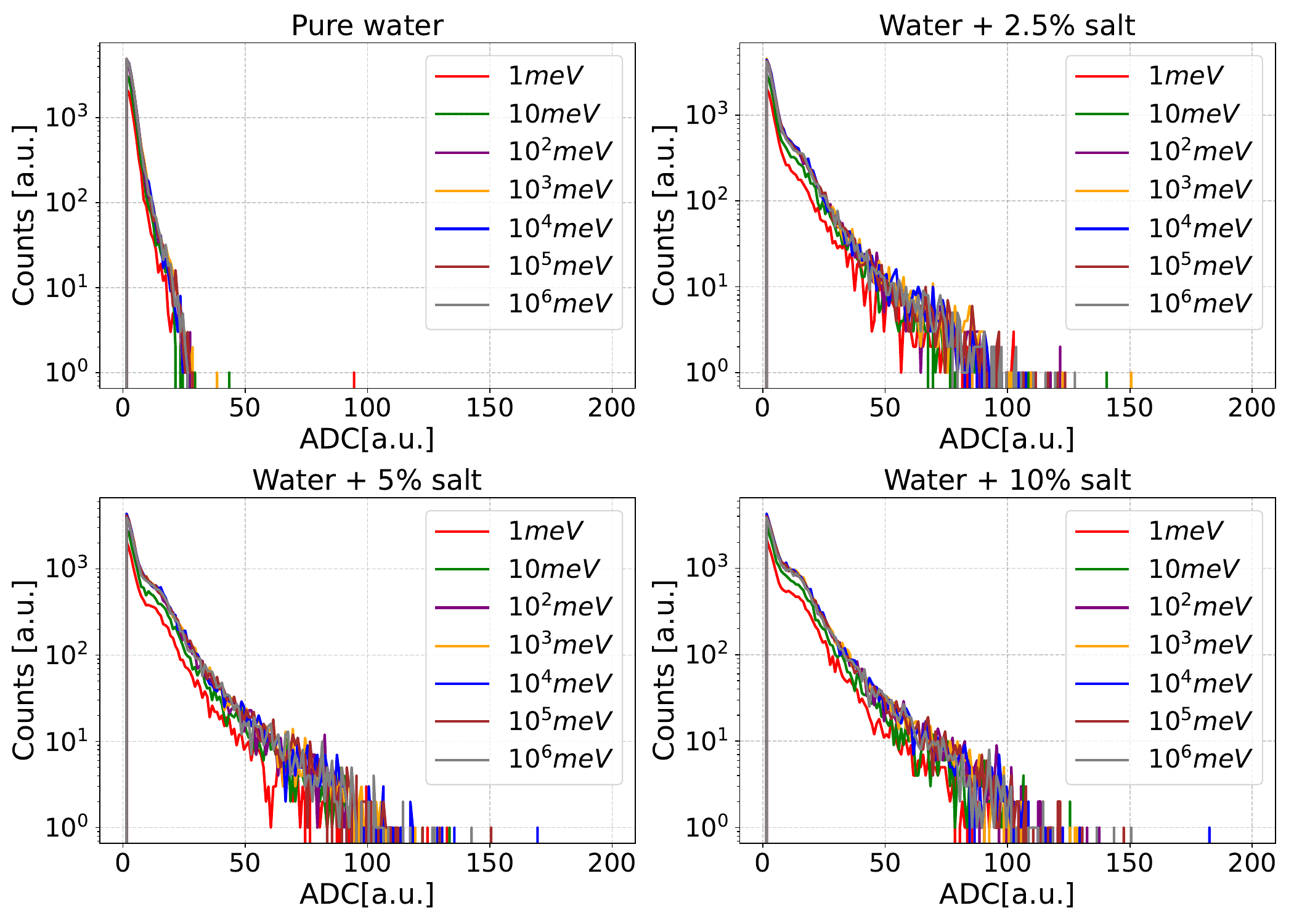}
\end{center}
\caption{Histograms of the total charge registered by the water Cherenkov detector, representing the number of Cherenkov photons detected and accepted according to the photomultiplier tube quantum efficiency. The upper left panel shows the photon distribution in pure water for incident neutrons of different energies: 1~meV (\textcolor{red}{red}), $10$~meV (\textcolor{green}{green}), $10^2$~meV (\textcolor{violet}{violet}), $10^3$~meV (\textcolor{orange}{orange}), $10^4$~meV (\textcolor{blue}{blue}), $10^5$~meV (\textcolor{brown}{brown}), and $10^6$~meV (\textcolor{gray}{grey}). The other three panels display the corresponding charge distributions for NaCl-doped water at increasing concentrations: 2.5\% (upper right), 5\% (lower left), and 10\% (lower right) by mass. The enhanced photon yield with increasing salt concentration reflects the higher neutron-capture cross section of $^{35}$Cl, which leads to the emission of additional high-energy gamma.}
\label{matriz-carga}
\end{figure}

\section{Results and discussion}
\label{ResultsDiscussion}
This section synthesises the numerical results of our MEIGA Geant4 study and translates them into practical design criteria for NaCl-doped water Cherenkov detectors. We first analyse thermal neutrons with $E_{n}<1$~keV, which dominate the ground-level flux and therefore set the detector trigger rate. We then evaluate the detector in an operational context by exposing it to the thermal component of the cosmic neutron background. This spectrum, covering $10^{-8}$~MeV $\leq  E_{n} \leq 10^{-5}$~MeV, is normalised to the atmospheric profile of Bucaramanga, Colombia.
\subsection{Evaluation for thermal regime}
We quantified the detector's response to thermal neutrons by tracking the capture efficiency, i.e
\[ \eta_{capt} =\frac{N_{capt}}{N_{inc}} \equiv \frac{\text{Number of captured neutrons}}{\text{Number of Incident neutrons}}\, ,
\]
where $N_{cap}$ is the number of neutrons absorbed in the active volume and $N_{inc}$ is the number of incident neutrons ( $N_{inc}=10^{5}$ in every simulation). We injected mono-isotropic neutron spectra uniformly distributed between $1$~meV and
$1$~keV. This interval covers the entire epithermal domain relevant to soil moisture sensing~\citep{KohliEtal2015}.

Figure~\ref{cap-det-term} (a) presents the capture efficiency as a function of neutron energy, $\eta_{capt}(E_n)$, for pure water and three NaCl concentrations. Across the entire thermal to epithermal range ($1$~meV~$\leftrightarrow~1$~keV), every salt concentration yields a higher efficiency than pure water. Chlorine 35, whose thermal cross section exceeds that of hydrogen by two orders of magnitude, drives this improvement~\citep{SidelnikEtal2020A}. Below $100$~meV, the \textcolor{orange}{orange} curve ($10$\% NaCl)  rises from $\eta_{capt} \approx 0.28$ to $>0.5$, whereas the \textcolor{red}{red} curve (pure water) lags near $0.18\lesssim~\eta_{capt}~\lesssim 0.35$.  All curves flatten around $\sim 1$eV, indicating that further increases in neutron energy no longer boost the capture probability. A light salt loading (ranging $2.5\% \leftrightarrow 5$\%) already secures most of the possible performance gain while incurring minimal cost and optical penalty, making such concentrations optimal for practical water Cherenkov neutron sensors.

Panel~(b) of the same Figure~\ref{cap-det-term}, displays the relative capture efficiency, which measures the gain introduced by NaCl doping. Every curve stays above unity throughout the ($1~\text{meV}\le E_n \le 1~\text{keV}$) range, so even the lightest admixture ($2.5$\% NaCl) increases the number of captured neutrons around $20$\%, for energies $\sim 1$meV. The large thermal cross section of $^{35}\mathrm{Cl}$ furnishes the additional absorption channels that drive this enhancement. From energies starting around $E_n \sim 1$meV the $\eta_{capt-rel}$ decreases. As the neutron energy approaches $1$keV, the ratios converge to $1.35$, $1.22$, and $1.12$ for $10$\%, $5$\%, and $2.5$\% NaCl, respectively. This convergence follows the canonical $1/v$ decline of the capture cross section and reproduces earlier Monte-Carlo predictions~\citep{SidelnikEtal2020B}. Because the $5$ \% and $10$ \% curves differ by less than $0.15$ across the spectrum, concentrations above $\sim 5$ \% deliver only marginal gains while imposing higher optical losses and material costs.

Neutrons that impinge on the water-air interface without being absorbed may scatter back into the atmosphere. Figure~\ref{cap-det-term} \textbf{(c)} presents the reflection coefficient $\eta_{\mathrm{refle}}(E_{n})$ for pure water and the same three NaCl mixtures. NaCl systematically lowers the reflected fraction: at $E_{n}\sim 1$meV pure water returns $\eta_{refle} \approx 0.55$, whereas a $10$\% solution yields $\eta_{refle}\approx 0.48$.  The contrast increases around $E_{n}~\sim~1$eV. Pure water still reflects nearly one neutron in two, but $10$\% NaCl restricts the fraction to $\eta_{refle}\sim 0.35$. All curves descend from the thermal domain to a broad minimum near the epithermal boundary, indicating that higher energy neutrons penetrate the interface more readily before being redirected by isotropic scattering. Because the $5$\% and $10$\% curves differ by less than $0.05$ across the spectrum, brine concentrations above $\sim 5$\% offer only marginal extra suppression while they raise the cost and diminish optical transparency. By reducing surface rebounding and simultaneously enhancing capture, NaCl increases the overall interaction rate, thereby sharpening the detector's sensitivity to environmental neutron fluxes~\citep{SidelnikEtal2020C}.

Finally, panel~(d) of Fig.~\ref{cap-det-term} reports the lateral-transport coefficient $\eta_{Trans}(E_n)$, defined as the fraction of neutrons that escape through the sidewalls without undergoing capture or surface reflection. Pure water leaks most strongly, with the coefficient peaking near $\eta_{Trans}\sim 0.27$ and declining monotonically to a plateau of $\eta_{Trans}\sim 0.18$ around one keV. Introducing NaCl curtails this loss across the entire spectrum. A $2.5$\% solution already lowers $\eta_{Trans}$ by roughly $10$ \%.  Richer chloride content not only shortens the neutron mean free path by the large capture cross section of $^{35}\text{Cl}$ ($\sigma_{n,\gamma}=43.8$ b), but also flattens the energy dependence, with $10$\% NaCl, the coefficient varies by less than $0.2$ over three decades in energy. GEANT4 simulations by \cite{SidelnikEtal2020C} reproduce this trend, showing that chloride nuclei accelerate capture and thereby diminish the residence time available for lateral escape. Consequently, moderate salinities ($\sim 5$\%) achieve nearly the full reduction in leakage while preserving optical transparency and controlling costs, a conclusion consistent with earlier experimental work on NaCl-doped water Cherenkov detectors.

Figure~\ref{matriz-carga} contrasts the Cherenkov photon yield for pure water and for water doped with NaCl. The upper left panel (pure water) shows a monotonic rise in yield from $1$~meV (\textcolor{red}{red} curve) to $1$keV (\textcolor{gray}{grey} curve). The remaining panels display the spectra for $2.5$\%, $5$\%, and $10$\% concentrations of NaCl. Adding salt consistently boosts the result, most strongly at higher neutron energies, yet the curves retain their shape. Thus, the doped WCD retains its characteristic energy dependence while producing a stronger signal as chlorine nuclei absorb more neutrons and release additional capture $\gamma$.

Salt-doped water Cherenkov detectors markedly strengthen thermal neutron measurements. The large capture cross-section of $^{35}\mathrm{Cl}$ boosts detection efficiency, while the denser medium suppresses both neutron reflection and lateral transport. These combined effects increase the Cherenkov photon yield and stabilise the counting rate, resulting in a stronger and more reliable signal. Such performance makes chloride-enhanced WCDs well-suited for ground-based monitoring of the thermal neutron field, providing an indirect yet sensitive proxy for volumetric soil water content —an essential variable in environmental and hydrological research.

\begin{figure}
    \centering
    \includegraphics[width=0.49\linewidth]{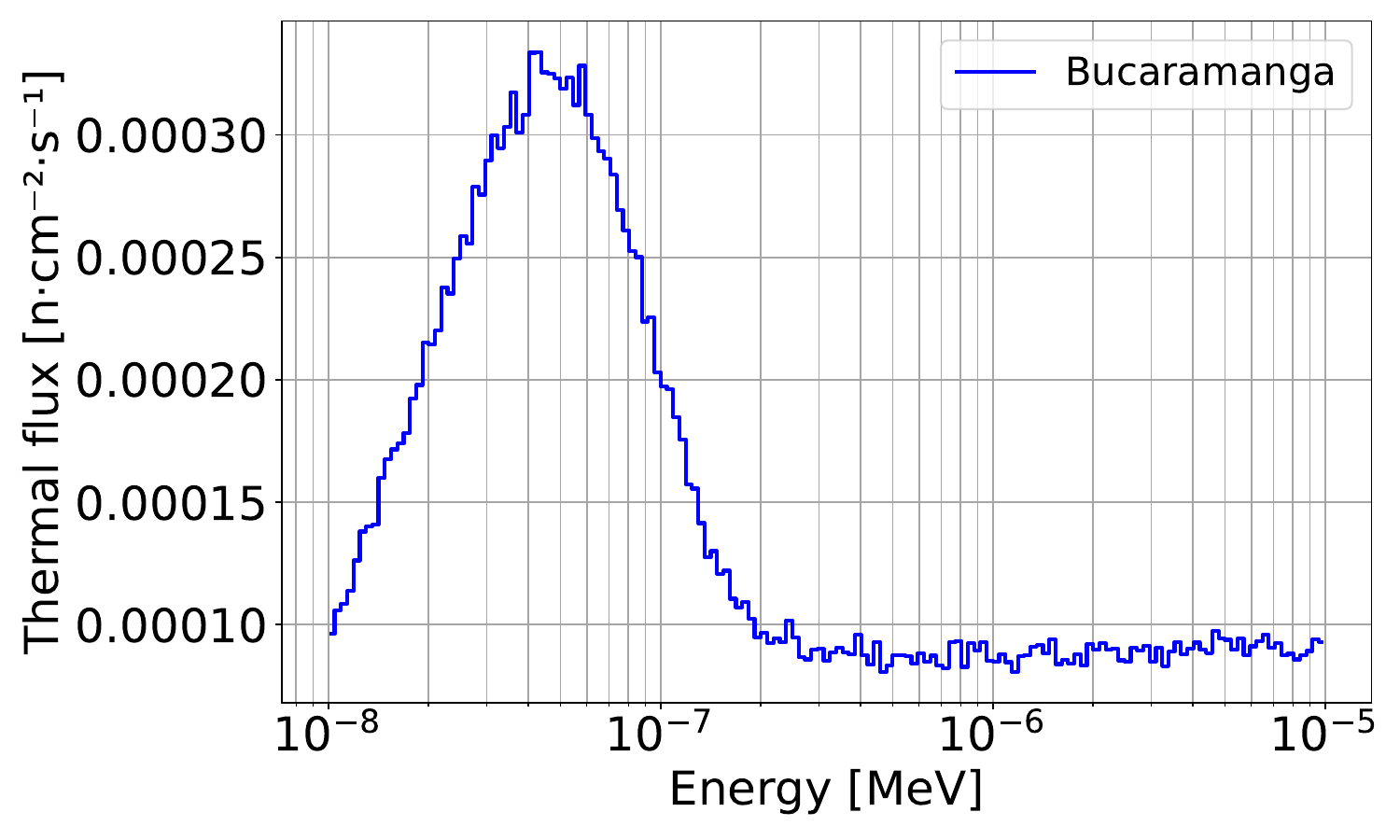}
    \includegraphics[width=0.49\linewidth]{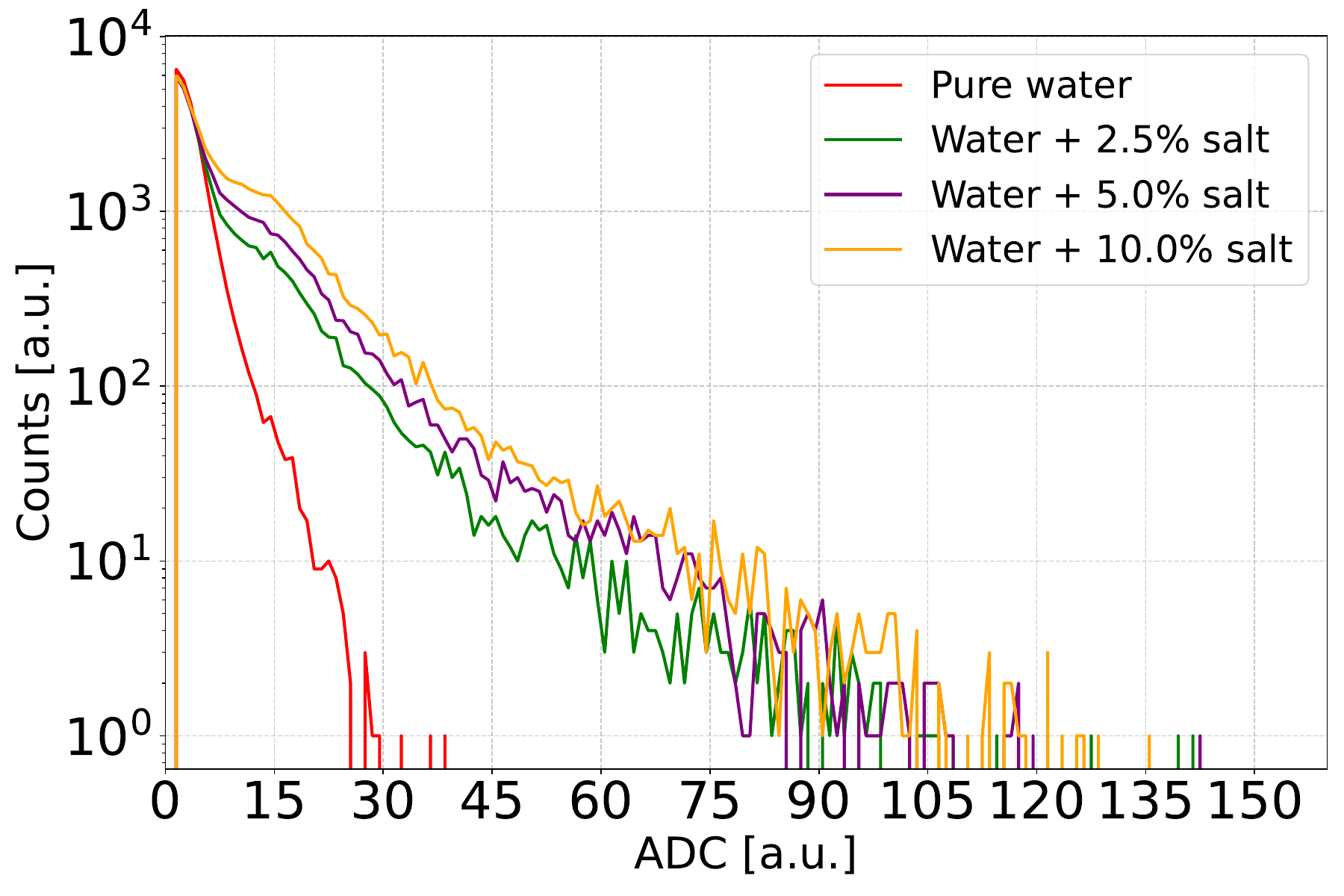}
    \caption{A 12-hour thermal neutron flux is initiated from a distance of 20 cm from the WCD, as shown on the left side of the figure. On the right, the charge response histograms for the four different configurations are displayed.}
    \label{flux-response}
\end{figure}

\subsection{Thermal neutron flux in Bucaramanga}
%%%%%%%%%
To evaluate performance under realistic field conditions, we modelled the detector's response to the ground-level thermal cosmic-neutron background. Using the procedure of Section \ref{subsec:neutron-flux}, we generated the relevant spectrum --$10$meV to $10$eV-- for the atmospheric profile of Bucaramanga, Colombia. We then injected this flux into the detector over a 12-hour exposure for four media: pure water and water doped with $2.5$\%, $5$\%, and $10$\% concentrations of NaCl.

Figure \ref{flux-response} displays the simulated thermal neutron spectrum (left) with the resulting Cherenkov photon output (right). In the Cherenkov Photon plot, two patterns emerge. The pure water curve (\textcolor{red}{red}) peaks sharply at 0-5 photons, revealing that many thermal neutrons generate only faint or subthreshold signals. Beyond five photons and up to $\sim 105$, the salt-doped media ($2.5$\%, $5$\%, and $10$\% NaCl; \textcolor{green}{green}, \textcolor{purple}{purple}, and \textcolor{orange}{orange}, respectively) diverge decisively from pure water: steeper slopes and higher counts testify to chloride-enhanced capture $\gamma$ and, therefore, a stronger detector response.

\begin{table*}[h]
    \centering
    \resizebox{\textwidth}{!}{%
    \begin{tabular}{|c|c|c|c|c|}
        \hline
        \multicolumn{5}{|c|}{\textbf{Thermal neutrons initiated 20 cm away from the water Cherenkov detector}} \\
        \hline
        \textbf{Type} & \textbf{Pure water} & \textbf{Water + 2.5\% salt} & \textbf{Water + 5\% salt} & \textbf{Water + 10\% salt} \\
        \hline
        Captured($\%$) & 23.5 & 27.1 & 30.2 & 35.2 \\
        Reflected($\%$) & 46.3 & 43.7 & 41.2 & 37.5 \\
        Transmitted($\%$) & 20.7 & 19.7 & 19.1 & 17.8 \\
        Did not enter($\%$) & 9.5 & 9.5 & 9.5 & 9.5 \\ 
        Incident($\%$) & 100 & 100 & 100 & 100 \\ 
        \hline
    \end{tabular}%
    }
    \caption{Performance metrics for $207,969$ simulated thermal neutrons --$10$\% of the 12 h ground level flux expected in Bucaramanga-- tracked in a water Cherenkov detector filled with pure water or water doped with $2.5$\%, $5$\%, or $10$\% NaCl. The detector either captures, reflects, or lets each neutron escape. The table lists the corresponding fractions. Increasing salinity raises the capture fraction by up to $11.7$ percentage points while lowering both reflection and escape losses, confirming the efficiency gains of chloride-enhanced media for thermal neutron sensing.}
    \label{porcentaje-neu}
\end{table*}

We simulate $207,969$ thermal neutrons (approximately $10$\% of the 12h ground-level flux expected in Bucaramanga) and track their trajectories in four media. As can be appreciated from the data displayed in the Table \ref{porcentaje-neu}, raising the NaCl content consistently improves detector performance. The capture fraction rises from $23.5$\% (pure water) to $35.2$\% ($10$\% NaCl), a gain of $11.7$ percentage points. Neutrons reflected at the water surface drop from $46.3$\% to $37.5$\%. The fraction that exists without interaction plunges from $20.7$\% to $2.9$\%. Higher salinity, therefore, suppresses neutron losses and boosts signal-producing captures, confirming the efficacy of chloride-enhanced water Cherenkov detectors for thermal neutron sensing.

Doping the detector water with NaCl radically transforms the capture scenario. In pure water, most thermal neutrons strike hydrogen, with oxygen providing a minor secondary sink. Each event yields a single capture $\gamma$. Adding salt inserts three efficient absorbers ($^{35}\mathrm{Cl}$, $^{37}\mathrm{Cl}$, and $^{23}\mathrm{Na}$) whose large thermal cross sections outcompete both hydrogen and oxygen. These extra channels lift the overall capture probability and multiply the number of emitted $\gamma$ inside the active volume, thereby amplifying the Cherenkov signal. This is clear from the data displayed in Table \ref{comp-neu-gamma}. Chlorine nuclei possess far larger thermal-capture cross sections than hydrogen.  When the salinity reaches $10$\%, captures on $^{35,37}\mathrm{Cl}$ surge to $68$\% of the total, while captures on $^1\mathrm{H}$ drop to $31$\%. The absolute number of captures rises by $50$\% ($48,825 \rightarrow 73,157$). A chlorine capture emits an $\sim 8.6$~MeV $\gamma$ cascade, whereas a hydrogen capture releases a single $2.2$~MeV photon. The higher multiplicity and energy boost the per capture Cherenkov yield; the WCD, therefore, registers $3.4$ times more $\gamma$ particles and $1.9$ times more Cherenkov inducing interactions at $10$\% NaCl. The intensified, harder photon field enhances the signal-to-noise ratio. These findings corroborate earlier prototype work on chloride-enhanced water Cherenkov detectors and show that a modest ($\leq 10$\% w/w) NaCl addition offers a cost-effective route to elevate detector sensitivity for environmental neutron measurements substantially.

%%%%%%%%%

\begin{table*}[h]
    \centering
    \scriptsize
    \resizebox{\textwidth}{!}{%
    \begin{tabular}{|c|ccc|ccc|ccc|ccc|}
        \hline
        \multicolumn{13}{|c|}{\textbf{207969 thermal neutrons sent to the water Cherenkov detector.}} \\
        \hline
        \textbf{Isotope} 
        & \multicolumn{3}{c|}{\textbf{Pure water}} 
        & \multicolumn{3}{c|}{\textbf{Water + 2.5 $\%$ salt}} 
        & \multicolumn{3}{c|}{\textbf{Water + 5 $\%$ salt}} 
        & \multicolumn{3}{c|}{\textbf{Water + 10 $\%$ salt}} \\

        Rea./Emi. & \textbf{Capt.} & $\boldsymbol{\gamma}$ & \textbf{Cha-i}
                  & \textbf{Capt.} & $\boldsymbol{\gamma}$ & \textbf{Cha-i}
                  & \textbf{Capt.} & $\boldsymbol{\gamma}$ & \textbf{Cha-i}
                  & \textbf{Capt.} & $\boldsymbol{\gamma}$ & \textbf{Cha-i} \\
        \hline
    $\sideset{^{1}}{}{\mathop{H}} (n,\gamma)^{2}\!H$ & 48797 & 48797 &  & 35787 & 35787 &  & 29787 & 29787 &  & 22605 & 22605 &  \\

    $\sideset{^{35}}{}{\mathop{Cl}} (n,\gamma)^{36}\!Cl$ &  &  &  & 20233 & 56582 &  & 32280 & 90234 &  & 49618 & 138187 &  \\

    $\sideset{^{23}}{}{\mathop{Na}} (n,\gamma)^{24}\!Na$ &  &  &  & 319 & 1152 &  & 553 & 1995 &  & 777 & 2835 &  \\

    $\sideset{^{37}}{}{\mathop{Cl}} (n,\gamma)^{38}\!Cl$ &  &  &  & 47 & 125 &  & 104 & 263 &  & 149 & 420 &  \\

    $\sideset{^{16}}{}{\mathop{O}} (n,\gamma)^{17}\!O$ & 28 & 92 &  & 15 & 46 &  & 13 & 39 &  & 8 & 21 &  \\

    \textbf{Total} & 48825 & 48889 & 23336 & 56401 & 93692 & 30722 & 62737 & 122318 & 36041 & 73157 & 164068 & 44021 \\
    \hline
    \end{tabular}
    } % Fin del resizebox
    \caption{Gamma photon emission resulting from thermal neutron capture reactions in different detector media. Each triplet of columns corresponds to the number of neutrons captured, gamma photons emitted, and charge integral.}
    \label{comp-neu-gamma}
\end{table*}

\section{Conclusions}
\label{Conclusions}
We developed and validated an end-to-end Monte Carlo simulation chain that integrates the ARTI atmospheric cascade toolkit with the MEIGA+Geant4 detector model. The framework reproduces the $500$~MeV benchmark of \cite{SidelnikEtal2020C, SidelnikEtal2020A, SidelnikEtal2020B}  within statistical uncertainties, confirming that it captures all critical processes --from cosmic ray shower propagation through neutron moderation to the final photoelectron signal-- required for reliable performance studies of water Cherenkov detectors.
\vspace{0.3em}

The addition of NaCl introduces multiple new thermal neutron capture channels ($^{35,37}\mathrm{Cl}$ and $^{23}\mathrm{Na}$), with the thermal neutron absorption cross section of $^{35}\mathrm{Cl}$ exceeding that of hydrogen by up to two orders of magnitude. A 10\% NaCl solution increases the overall capture probability by approximately 50\%, shifting the dominant neutron absorber from hydrogen (in pure water) to chlorine, which accounts for 68\% of all captures in this configuration.

Also, neutron capture on chlorine yields a high-energy gamma cascade of approximately $\sim 8.6$~MeV, roughly four times the energy of the $2.2$~MeV photon released from hydrogen capture ($^1\mathrm{H}(n,\gamma)$ and $^2\mathrm{H}(n,\gamma)$). As a result, the total gamma energy deposited and the number of Compton electrons capable of producing Cherenkov light increase by factors of 3.4 and 1.9, respectively, in the 10\% NaCl solution. These enhancements lead to a significantly improved signal-to-noise ratio in the detector response.

On the other hand, we observe that a 5\% NaCl admixture already achieves over 80\% of the maximum performance enhancement, while maintaining low optical attenuation and material costs. Beyond this concentration, further salt addition yields diminishing returns, as increased photon absorption in the brine begins to offset the nuclear gains.

When a realistic 12-hour background neutron flux scenario for Bucaramanga was probed, the salt-doped detector doubled the number of events above the five-photon threshold. It reduces reflection and escape losses from 67\% (in pure water) to 40\%. These improvements enhance the correlation between detector count rate and soil moisture content, reinforcing the applicability of cosmic ray neutron sensing techniques for environmental monitoring and precision agriculture.

Chloride-enhanced WCDs provide a non-toxic and inexpensive route to enhance thermal neutron detection. Their hectare-scale footprint and robustness make them attractive complements to existing CRNS networks for hydrological and precision agriculture applications. Field campaigns will calibrate the simulated gains against in situ moisture probes, assess the long-term optical stability of brine solutions, and explore combined epithermal and thermal readouts. Integrating low-power electronics and remote telemetry will further pave the way for large-area deployments in varied climatic settings.

\section*{Acknowledgements}
JB gratefully acknowledges the support of the Colombian Ministry of Science, Technology, and Innovation (MinCiencias), as this work was made possible through funding from the Bicentennial Scholarship Program, second call. JB also extends sincere thanks to the University of Nariño for its financial support throughout the entire academic training conducted at the School of Physics of the Universidad Industrial de Santander. This research also benefited from the support of the ERASMUS+ internship program within the framework of the ``Latin American Alliance for Capacity Building in Advanced Physics'' (LA-CoNGA physics) project, during which the initial calculations that led to this study were conducted. The authors further acknowledge the co-funding provided by CYTED through the LAGO-INDICA network (Project 524RT0159-LAGO-INDICA: Infraestructura digital de ciencia abierta). AI technology was used to proofread and polish this manuscript. (OpenAI, 2025)

\bibliographystyle{elsarticle-harv}  
\bibliography{NeutronCherenkov}

\end{document}